\documentclass[prl,twocolumn]{revtex4}
\usepackage{dcolumn}
\usepackage{multirow}
\usepackage{graphicx}
\usepackage{amssymb}
\usepackage{bm}
\usepackage{hyperref}
\usepackage{epstopdf}
\usepackage{color}
\usepackage{mathrsfs}
\usepackage{amsmath,amssymb,amsthm}
\usepackage{rotating}
\usepackage{sverb, longtable}
\usepackage{subfigure}

\usepackage[graphicx]{realboxes}
\usepackage{adjustbox}

\begin{document}

\title{Inflation 2024}
\author{Deng Wang}
\email{dengwang@ific.uv.es}
\affiliation{Instituto de F\'{i}sica Corpuscular (CSIC-Universitat de Val\`{e}ncia), E-46980 Paterna, Spain}

\begin{abstract}
Inflation as the leading paradigm depicting the very early universe physics could leave imprints on the cosmic microwave background (CMB) radiation. Using currently available CMB observations, we give the tightest constraints on inflation so far. We discuss the theoretical implications of our results including the energy scale of inflation, inflaton field excursion, Hubble expansion rate during inflation, equation of state of inflation, primordial tensor non-Gaussianity, primordial tensor power spectrum, B-mode anisotropy and inflationary gravitational wave background.
\end{abstract}


\maketitle

{\it Introduction.} The biggest question in science is the origin of the universe. In the hot big bang paradigm, shortly after the singularity, the universe is assumed to undergo an quasi-exponential expansion phase around $10^{-34}$ s, i.e., the so-called inflation \cite{Brout:1977ix,Starobinsky:1980te,Kazanas:1980tx,Sato:1981qmu,Guth:1980zm,Linde:1981mu,Linde:1983gd,Albrecht:1982wi}, which sets the very special initial conditions of such a huge universe we observe today. After inflation, the universe experiences in succession the (p)reheating, radiation-dominated, recombination, dark ages, reionization and dark energy eras. Clearly, as the onset of the universe, inflation will leave imprints on all the observations from a later stage. Although the nature of inflation is still unknown, we can probe the earliest moments of the universe now, due to well theoretical modeling and high precision observations \cite{Martin:2013tda,Achucarro:2022qrl,Baumann:2009ds}. To decode the physics of inflation, the main tool is the statistics of primordial scalar and tensor fluctuations, where the former seeds the formation of large scale structure (LSS) and induces the primordial features \cite{Slosar:2019gvt,Chluba:2015bqa} and primordial non-Gaussianity (PNG) \cite{Maldacena:2002vr,Meerburg:2019qqi,Chen:2010xka,Collins:2011mz}, while the latter produces the primordial gravitational waves (PGWs) \cite{Achucarro:2022qrl,Seljak:1996gy,Krauss:2010ty,Kamionkowski:2015yta,Caprini:2018mtu} as the earliest ripples of spacetime. These three important predictions of inflation help extract physical information at different inflationary energy scales and could be detected by the present-day CMB and LSS observations \cite{Krauss:2010ty}. Up to now, the Planck CMB data \cite{Planck:2019nip,Planck:2018lbu} has constrained the deviation from the almost scale-invariant primordial scalar spectrum to be less than one percent of the amplitude $A_s$ \cite{Planck:2018vyg,Planck:2018jri} and shown no evidence of PNG \cite{Planck:2019kim}. Very interestingly, using a logarithmic prior of the tensor-to-scalar ratio $r$, we have recently given the strongest constraints on PGWs via three available CMB experiments including the Planck \cite{Planck:2018vyg,Planck:2018jri}, Atacama Cosmology Telescope (ACT) \cite{ACT:2020frw,ACT:2020gnv} and the South Pole Telescope (SPT) \cite{SPT-3G:2021eoc,SPT-3G:2022hvq} at the $2\,\sigma$ confidence level (CL), and provided the corresponding $2\,\sigma$ upper bounds on $\log_{10}r$ \cite{Wang:2024hhh}, which have outstanding implications on the very early universe physics.        

In this work, we present the new constraints on PGWs from the BICEP/Keck Array CMB B-mode polarization data \cite{BICEP:2021xfz} and novel possibilities from our data analysis and, more importantly, interpret the theoretical implications of our constraining results of PGWs in the framework of the single-field slow-roll inflation, including the energy scale of inflation, inflaton field excursion, Hubble expansion rate during inflation, inflationary equation of state (EoS), primordial tensor non-Gaussianity, primordial tensor power spectrum (PTPS), angular power spectrum of CMB B-modes and inflationary gravitational wave background (IGWB). 

{\it New limits and novel possibilities.} Following the analysis methodology of the Planck collaboration \cite{Planck:2018vyg,Planck:2018jri} and our previous work \cite{Wang:2024hhh}, we write the PTPS as
\begin{equation}
\mathcal{P}_{T}(k)=A_t\left(\frac{k}{k_p}\right)^{n_t+\frac{1}{2}n_{trun}\ln\left(\frac{k}{k_p}\right)}, \label{eq:PTS}
\end{equation}  
where $k$, $k_p$, $A_t$, $n_t$ and $n_{trun}$ are, respectively, the comoving wavenumber, tensor pivot scale, amplitude of PTPS, tensor spectral index and running of tensor spectral index. $r\equiv A_t/A_s$, where $A_s$ denotes the amplitude of primordial scalar power spectrum. The next-order inflation consistency relation (ICR) employed by the Planck collaboration, namely $n_t=-r(2-r/8-n_s)/8$ and $n_{trun}=r(r/8+n_s-1)/8$ \cite{Planck:2018vyg,Planck:2018jri,Planck:2015sxf} is used in this work, where $n_s$ is the scalar spectral index.

\begin{figure*}
	\centering
	\includegraphics[scale=0.5]{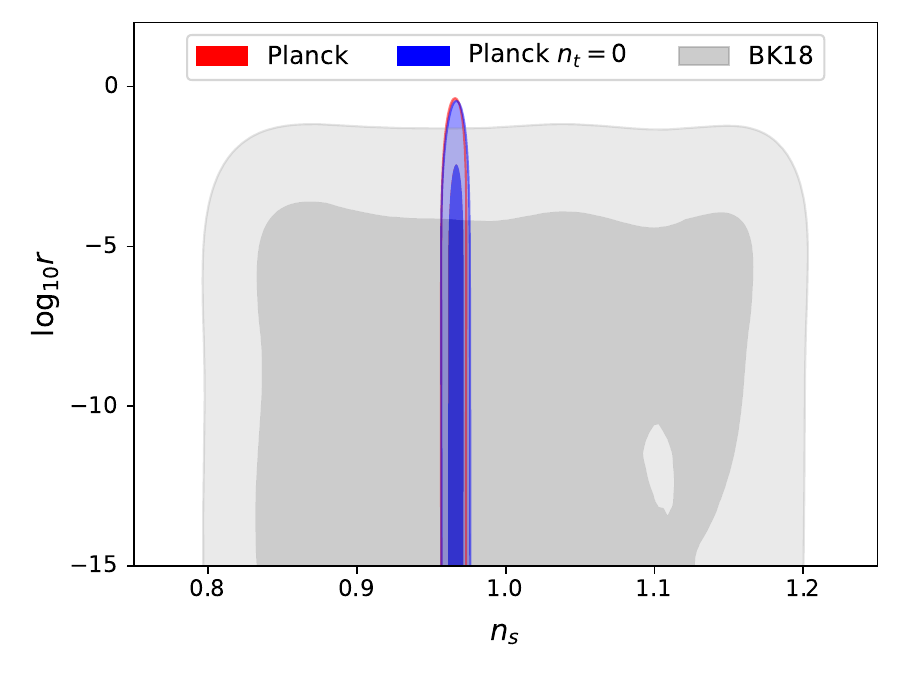}
	\includegraphics[scale=0.5]{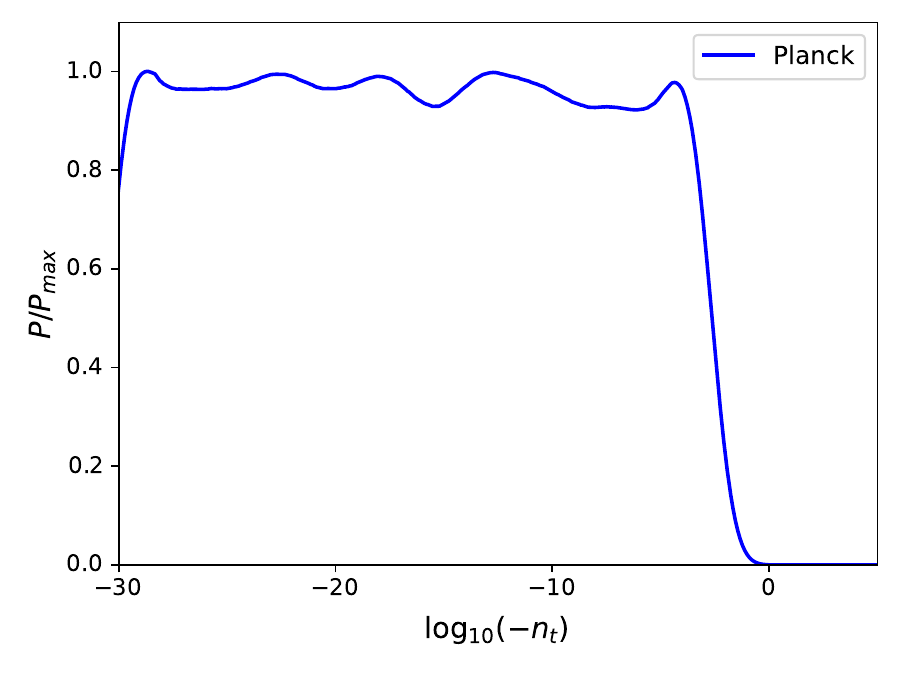}
	\caption{{\it Left.} The two-dimensional marginalized posterior distributions of the parameter pair ($n_s$, $\log_{10}r$) from the Planck and BK18 observations. We consider the cases of next-order ICR and $n_t=0$ when using the Planck data. {\it Right.} The derived constraint on the tensor spectral index $n_t$ from the Planck data for a single-field slow-roll inflation with the next-leading-order ICR.}\label{fig:pgw}
	
\end{figure*}

\begin{figure}
	\centering
	\includegraphics[scale=0.28]{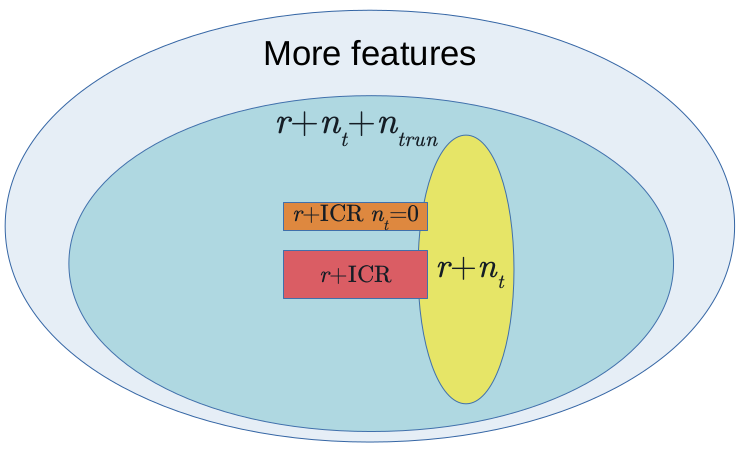}
	\caption{The schematic plot for the relations between a single-field slow-roll inflation with the next-order ICR (red) or with next-order ICR but $n_t=0$ (orange) and primordial tensor features including $\Lambda$CDM+$r$+$n_t$, $\Lambda$CDM+$r$+$n_t$+$n_{trun}$ and extended tensor features.}\label{fig:fts}
\end{figure}

To implement the Bayesian analysis, we adopt the following two datasets: (i) Planck. The Planck 2018 high-$\ell$ \texttt{plik} temperature (TT) data at multipoles $30\leqslant\ell\leqslant2508$, polarization (EE) and their cross-correlation (TE) likelihoods at $30\leqslant\ell\leqslant1996$, and the low-$\ell$ TT \texttt{Commander} and \texttt{SimAll} EE data at $2\leqslant\ell\leqslant29$ \cite{Planck:2019nip} as well as the Planck lensing likelihood \cite{Planck:2018lbu} from the \texttt{SMICA} maps at $8\leqslant\ell \leqslant400$. (ii) BK18. The BICEP2, Keck Array and BICEP3 CMB polarization data up to the 2018 observing season \cite{BICEP:2021xfz}. This dataset aiming at detecting the CMB B-modes contains the additional Keck Array observations at 220 GHz and BICEP3 observations at 95 GHz relative to the previous 95/150/220 GHz data. 

Furthermore, we take the public Boltzmann code \texttt{CAMB} \cite{Lewis:1999bs} to compute the theoretically cosmological quantities and adopt the Monte Carlo Markov Chain (MCMC) method to infer the posterior probability density distributions of model parameters using the package \texttt{CosmoMC} \cite{Lewis:2002ah,Lewis:2013hha}. The generated MCMC chains is post-processed by the package \texttt{Getdist} \cite{Lewis:2019xzd} and the corresponding convergence rule of each MCMC run is the Gelman-Rubin diagnostic $R-1\lesssim 0.02$ \cite{Gelman:1992zz}. Same as our previous analysis \cite{Wang:2024hhh}, the following uniform priors for seven model parameters are used: the the logarithmic tensor-to-scalar ratio $\log_{10}r \in [-30, 5]$,  $\mathrm{ln}(10^{10}A_s) \in [2, 4]$, $n_s \in [0.8, 1.2]$, baryon fraction $\Omega_bh^2 \in [0.005, 0.1]$, cold dark matter fraction $\Omega_ch^2 \in [0.001, 0.99]$, comoving acoustic angular scale at the recombination epoch $100\theta_{MC} \in [0.5, 10]$ and optical depth due to reionization $\tau \in [0.01, 0.8]$. The pivot scale used is 0.05 Mpc$^{-1}$.

\begin{figure*}
	\centering
	\includegraphics[scale=0.43]{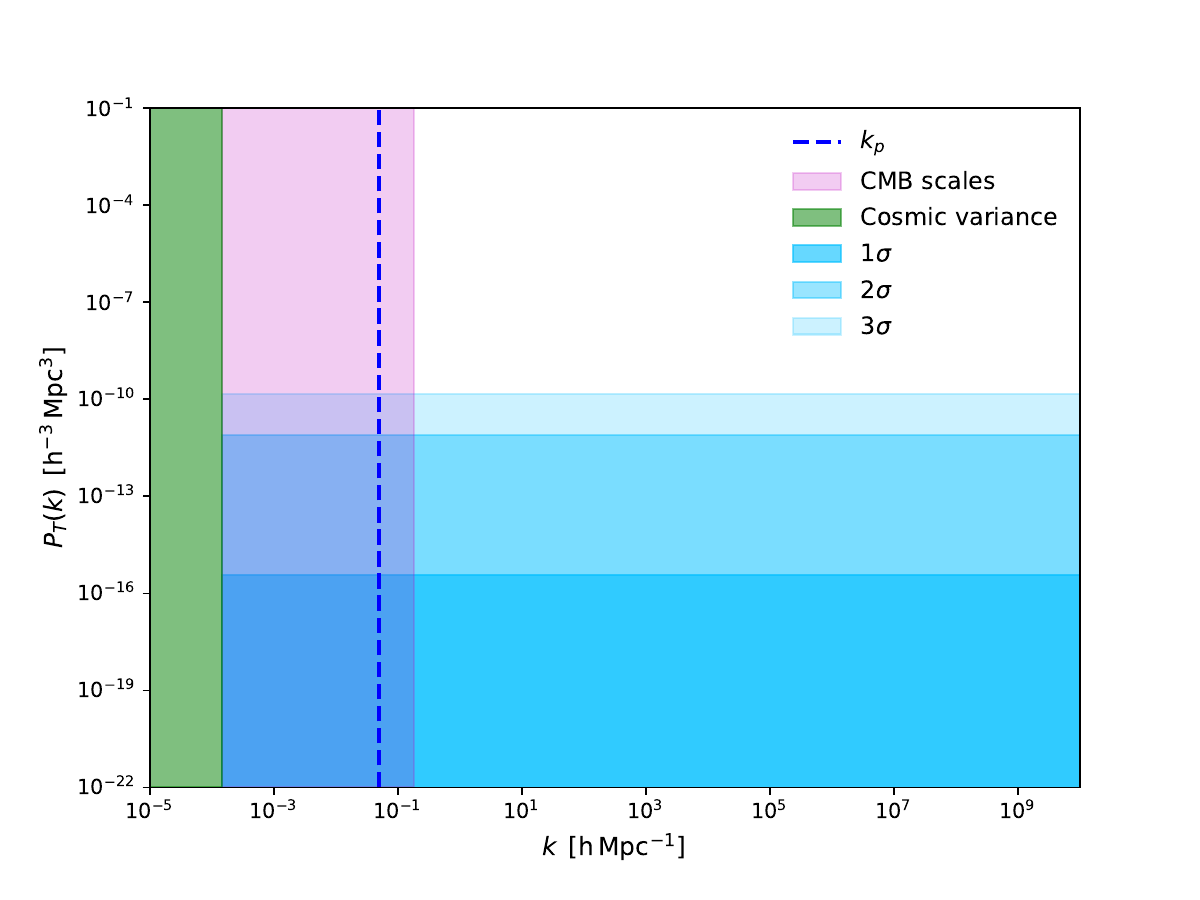}
	\includegraphics[scale=0.43]{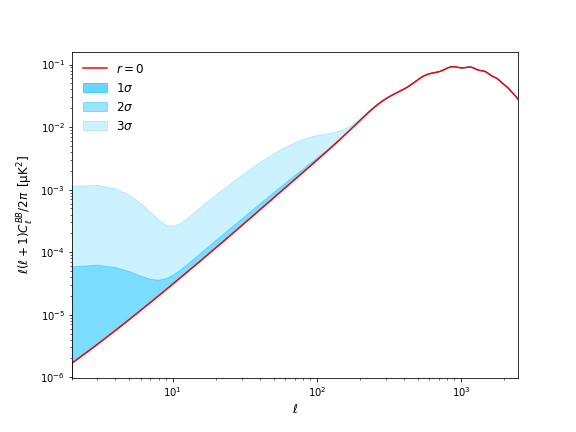}
	\caption{The constrained primordial tensor power spectrum ({\it left}) and lensed CMB B-mode angular power spectrum ({\it right}) with $1\,\sigma$, $2\,\sigma$ and $3\,\sigma$ confidence regions. The magenta and green shaded regions denote the CMB scales and the cosmic variance, respectively. The red solid and blue dashed lines represent the case of $r=0$ and the pivot scale $k_p=0.05$ Mpc$^{-1}$, respectively. }\label{fig:pts}
\end{figure*}

Our previous results \cite{Wang:2024hhh} have presented the strongest constraints from three CMB experiments including Planck, ACT and SPT separately as well as their combinations with BK18 and DESI observations \cite{DESI:2024mwx,DESI:2024uvr,DESI:2024lzq} at the $2\,\sigma$ CL. However, the constraint on six free parameters of $\Lambda$CDM plus $\log_{10}r$ from BK18 alone is not investigated, even though it can not give any restriction on the scalar spectral index. Instead, we just constrain $\log_{10}r$ by assuming the Planck 2018 fiducial cosmology. In this study, we show the BK18 constraint on $\Lambda$CDM+$\log_{10}r$ in Fig.\ref{fig:pgw}, which gives $\log_{10}r<-2.472$ at the $2\,\sigma$ CL. Actually, this $2\,\sigma$ upper bound is slightly tighter than our previous constraint $\log_{10}r<-2.433$ from the Planck data \cite{Wang:2024hhh}. It is noteworthy that all the above results are obtained by assuming the second-order ICR in a single-field slow-roll inflation. Interestingly, there is a novel possibility that PGWs can still exist when fixing $n_t=0$. The reason why we consider this case is that current cosmological data only provides a very weak constraint on free $n_t$, which does not rule out $n_t=0$ \cite{Planck:2019kim}. Although $n_t=0$ is different from the theoretical prediction of inflation $n_t<0$, it has interesting physical properties, namely reducing the primordial tensor parameter space and only introducing high order scale-dependence of PTPS. In light of the Planck data, we find that the constraints for the case of $n_t=0$ are completely same as the single-field inflation with the full ICR.

As mentioned above, one of the most important prediction is a negative $n_t$. Consequently, we study this point in a single-field slow-roll inflation with next-order ICR using the Planck observations, and derive the bound ${\rm log}_{10}(-n_t) < -3.323$ at the $2\,\sigma$ CL. In Fig.\ref{fig:pgw}, one can easily find that $n_t<0$ is well consistent with the prediction of inflation. However, we can not conclude that inflation is found at the current stage, since $r=0$ can not be ruled out by data. 

So far, the constraints on PGWs depends strongly on the next-order ICR. But, we do not explore whether PGWs exist in extended parameter spaces such as $\Lambda$CDM+$r$+$n_t$, $\Lambda$CDM+$r$+$n_t$+$n_{trun}$ or more tensor features. We show roughly the relations between models considered here and possibly extended features in Fig.\ref{fig:fts}. The reason why there is an overlap between $r$+ICR and $r$+$n_t$ is that the posterior of $n_t$ in $r$+ICR is a small subset of that in $r$+$n_t$ and the posterior of $n_{trun}$ in $r$+ICR includes $n_{trun}=0$ in $r$+$n_t$. Naturally, $r$+ICR with $n_t=0$ and $r$+$n_t$ display a similar overlap. The only difference is that $n_t$ lives in a small negative range satisfying the theoretical prediction of inflation in $r$+ICR relative to $r$+ICR with $n_t=0$. This is the reason why they have no any overlap. Since PGWs have been constrained to a high precision by currently independent CMB facilities \cite{Planck:2018vyg,Planck:2018jri,ACT:2020frw,ACT:2020gnv,SPT-3G:2021eoc,SPT-3G:2022hvq,BICEP:2021xfz,WMAP:2012nax}, it is very necessary to probe completely the primordial tensor parameter space using the same methodology and datasets depicted here. An in-depth exploration of whether there are signals of primordial tensor features is forthcoming \cite{Wang:2024kkk}.  

{\it Energy scale.} The detection of PGWs can provide a unique way to estimate the energy scale of inflation, which is expressed as $V_p=3\pi^2A_s/2rM_{\rm pl}^4=1.01\times10^{16} {\rm GeV}(r/0.01)^{1/4}$ \cite{Planck:2018jri} ($M_{\rm pl}$ is the reduced Planck mass) at the horizon exit of a pivot scale, since they carry the information of their generating mechanism. Using the constraints on $\log_{10}r$ from Planck \cite{Wang:2024hhh}, we obtain the inflationary energy scale as 
$V_p^{1/4}<7.91\times10^{15}$ GeV at the $2\,\sigma$ CL and $V_p^{1/4}<1.66\times10^{16}$ GeV at the $3\,\sigma$ CL, respectively. Planck's $2\,\sigma$ bound indicates that the energy scale of inflation is lower than the Grand Unified Theory (GUT) scale $\sim10^{16}$ GeV. Interestingly, its $3\,\sigma$ upper limit translated to $1.66\times10^{16}$ GeV is higher than the GUT scale. This implies that the electromagnetic, weak and strong forces could be unified during inflation. Although the energy scale is lower than the Planck scale $\sim10^{19}$ GeV, it is far larger than $\sim10^{4}$ GeV that can be reached by the most powerful high-energy particle facility on earth, i.e., the Large Hadron Collider (LHC). More intriguingly, our findings reveals that the physics before inflation should be more close to the Planck scale.

{\it Hubble expansion rate.} Using the above constraints on the energy scale of inflation and the Friedman equation $H_p^2\approx V_p/(3M_{\rm pl}^2)$, the Hubble parameter during inflation can be constrained to be $H_p/M_{\rm pl}<6.09\times10^{-6}$ at the $2\,\sigma$ CL and $H_p/M_{\rm pl}<2.67\times10^{-5}$ at the $3\,\sigma$ CL. The $2\,\sigma$ upper bound of $6.09\times10^{-6}$ is tighter than $2.5\times10^{-5}$ obtained by the Planck collaboration \cite{Planck:2018jri}.

\begin{figure}
	\centering
	\includegraphics[scale=0.43]{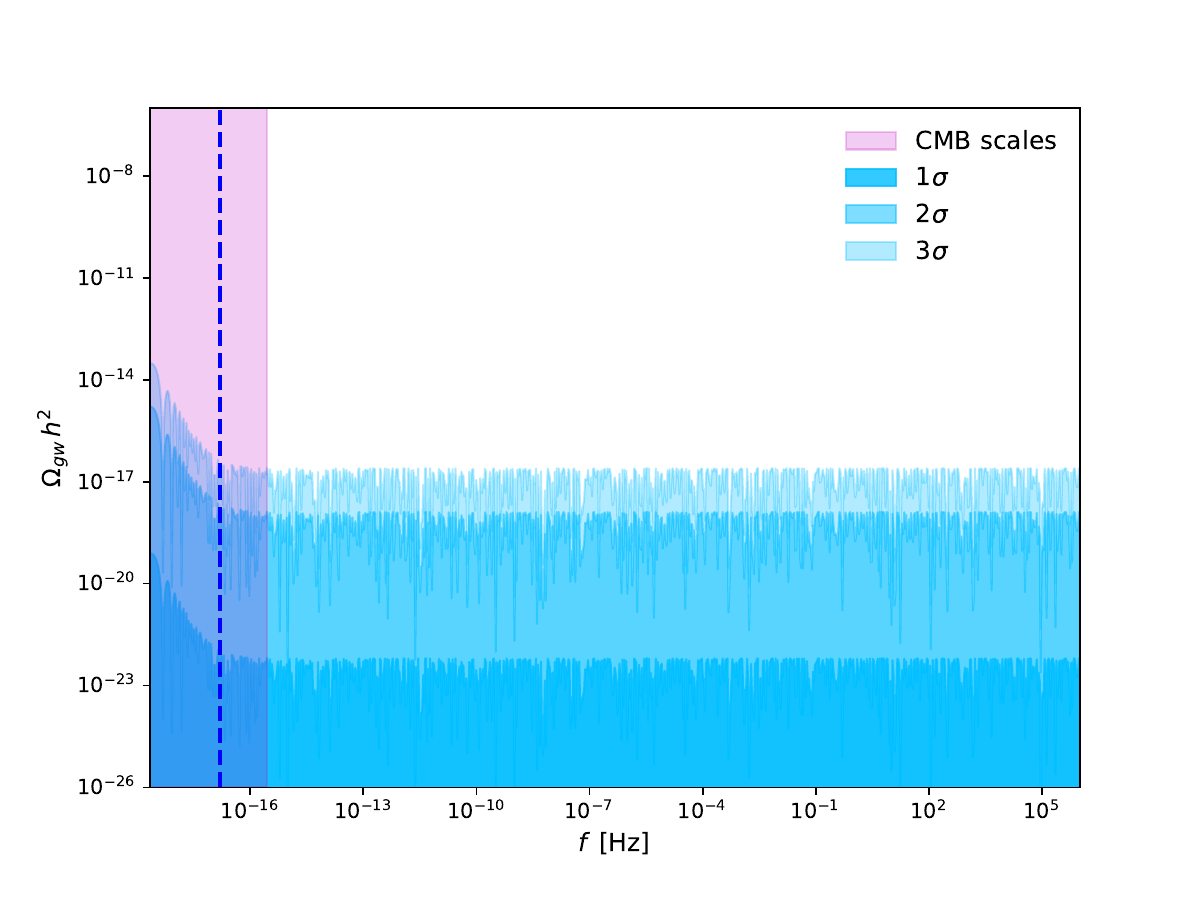}
	\caption{The constrained inflationary gravitational wave background with $1\,\sigma$, $2\,\sigma$ and $3\,\sigma$ confidence regions. The magenta shaded region and blue dashed line denote the CMB scales and the pivot scale $k_p=0.05$ Mpc$^{-1}$, respectively. }\label{fig:igwb}
\end{figure}

{\it Inflaton excursion.} The scalar field excursion $\Delta\phi$ characterizing the variation of inflaton field between the horizon exit of the pivot scale and the end of inflation provides the information about the viable quantum gravity theory. It is related to the tensor-to-scalar ratio via $\Delta\phi/M_{\rm pl}=(r/8)^{1/2}N_e$ \cite{Lyth:1996im}, where $N_e$ is the e-folding number during this period. The constraint of this field net excursion leads to a classification of larger-field and small-field inflationary models, where the discriminating value is the Planck mass. Choosing $N_e=30$ \cite{Boubekeur:2005zm,Boubekeur:2012xn}, Planck data gives the $2\,\sigma$ and $3\,\sigma$ upper bounds $\Delta\phi/M_{\rm pl}<0.644$ and $2.828$, respectively. One can easily find that the result is consistent with effective field theories supporting small-field excursions at the $2\,\sigma$ CL after removing the higher-order corrections by introducing some symmetries, and that the $3\,\sigma$ upper bound of $\Delta\phi/M_{\rm pl}=2.828$ from Planck is higher than the Planck scale implying a possible existence of super-Planckian field excursion, which poses a great challenge for theoretical modeling. Interestingly, our constraint is compatible with the first criterion of the swampland conjecture (SC) \cite{Ooguri:2006in,Obied:2018sgi,Agrawal:2018own}, i.e., $|\Delta\phi/M_{\rm pl}|<\Delta\sim\mathcal{O}(1)$ with $\Delta$ being a constant at the $2\,\sigma$ CL. However, it can place a strong constraint on the constant $c$ of the second criterion of SC, $M_{\rm pl}V'/V>c\sim\mathcal{O}(1)$, and gives $c<0.021\ll1$ at $2\,\sigma$ level using the relation $r>8c^2$ \cite{Matsui:2018bsy,Ben-Dayan:2018mhe}. Clearly, this bound poses a very strong tension with SC indicating that string-based inflationary models are likely not viable \cite{Wang:2018duq}.

{\it Nature of inflation.} The tensor spectral index can act as a function of inflation EoS $\omega$ via the formula $n_t=4/(1+3\omega)+2$ \cite{Mukhanov:1990me,Lyth:1998xn}. Using the Planck data, we obtain the $2\,\sigma$ upper limit of $\omega <-0.99984$ implying that inflation EoS is very close to $\omega=-1$, but we can not rule out $\omega=-1$ corresponding to $r=0$. This means that the nature of the single-field slow-roll inflation with the next-order ICR is possibly quintessence. Note that we only study the phenomenological nature of inflation here and do not consider its fundamental physics origin. As demonstrated in \cite{Wang:2024hhh}, cosmological $\alpha$-attractor inflation is well supported by observations. Interestingly, besides inflationary $\alpha$-attractors, there is still a large part of parameter space of $(n_s, \log_{10}r)$ needed to be explained by new physics.   

{\it Non-Gaussianity.} The theory predicts that primordial scalar and tensor non-Gaussianities have the forms of $f_{\rm NL}^s=5/12(n_s-1)$ and $f_{\rm NL}^t=5/24(3-n_t)$ after discarding the higher order terms \cite{Copeland:1994vg,Dine:1995uk,Maldacena:2002vr,Kundu:2013gha}. The Planck data has constrained $n_s$ to a $0.4\%$ precision and gives $f_{\rm NL}^s=-0.0146\pm0.0018$ at $1\,\sigma$ level \cite{Planck:2019kim}. Note that this constraint is obtained without inflationary model dependence. However, we just obtain a $2\,\sigma$ upper bound on the tensor non-Gaussianity $f_{\rm NL}^t<0.625$, which is derived in a single-field slow-roll inflation.    

{\it PTPS.} Since we have displayed strong constraints on the PTPS amplitude, we can give the observationally allowed PTPS across different scales. In Fig.\ref{fig:pts}, we present the constrained PTPS within the $3\,\sigma$ CL from Plank by using the next-order ICR. One can easily find that PTPS is scale-invariant due to very small $n_t$ and $n_{trun}$ and spans many orders of magnitude at $3\,\sigma$ level. This will lead to many underlying possibilities of the very early universe to interpret the observed PTPS. 

{\it B-mode anisotropy.} Similar to PTPS, in Fig.\ref{fig:pts}, we give the allowed angular power spectrum (APS) of CMB B-mode by Planck at $3\,\sigma$ level. Since B-mode mainly contributes to the CMB temperature APS at large scales, we study the B-mode APS at $2<\ell<200$. The constrained APS spans about three orders of magnitude within $3\,\sigma$ CL at the largest scale, say $\ell=2$, however, this large uncertainty span reduces to zero around $\ell=200$. 

{\it IGWB.} The stochastic gravitational wave background (SGWB) can be probed by pulsar timing arrays (PTA) \cite{Verbiest:2021kmt} and gravitational wave observatories \cite{LIGOScientific:2016jlg}. Recently, several PTA experiments such as NANOGrav \cite{NANOGrav:2023gor}, EPTA \cite{Antoniadis:2023rey}, PPTA \cite{Reardon:2023gzh} and CPTA \cite{Xu:2023wog} have reported the substantial evidences of SGWB. Interestingly, inflation can act as a possible cosmological source of SGWB. Although IGWB as a single source has been constrained by PTA data \cite{NANOGrav:2023hvm}, we argue that IGWB may not be the only source of SGWB. Specifically, we employ the IPTA DR2 \cite{Antoniadis:2022pcn} to constrain the single-field slow-roll inflation with next-order ICR and find ${\rm{log}_{10}}r = 0.485^{+0.013}_{-0.010}$ at $1\,\sigma$ level, which is much larger than the constraint from CMB data. Hence, we conclude that IGWB really exist in the universe but must be mixed with other astrophysical or cosmological sources to explain the SGWB. Moreover, considering appropriately the contribution of radiation-matter equality and reheating eras to the transfer function \cite{Guzzetti:2016mkm}, the gravitational wave energy spectrum constrained by Planck data is shown in Fig.\ref{fig:igwb}. Interestingly, when frequencies are larger than CMB ones, $\Omega_{\rm gw}h^2$ shows a clear frequency-invariant feature.       

{\it Concluding remarks.} The detection of PGWs plays a key role in understanding the connection between general relativity and quantum theory and probing the high energy physics close to the Planck scale. It can provide the direct evidence of the leading paradigm of the very early universe, i.e., inflation. By presenting new constraints and novel possibilities of PGWs, we compresses the life space of inflation that may possibly exists at the beginning of the universe in light of the fact that the Planck CMB experiments have given strong constraints on the PGWs.  

Furthermore, current data can not give a good constraint on the extended features such as $\Lambda$CDM+$n_t$+$n_{trun}$, and consequently, there is still a large tensor parameter space need to be explored by future observations and theoretical modeling. In the above, we just mention that $\alpha$-attractor inflation can explain the current constraints well. Actually, we require a large number of models to be proposed to interpret the large $n_s$-$\log_{10}r$ contour, where $r$ spans many orders of magnitude.

Since our constraining results have revealed that the GUT scale can be reached during inflation within $3\,\sigma$ level, the particle production of dark matter around the GUT scale could be allowed by observations. As a consequence, our constraints could inspire the exploration of dark matter model building at extremely high energy scales, and intriguingly, provide a new window to investigate the physics before inflation. 

{\it Acknowledgements.} DW is supported by the CDEIGENT Fellowship of Consejo Superior de Investigaciones Científicas (CSIC). DW warmly thanks the usage of Planck \cite{Planck:2018vyg}, ACT \cite{ACT:2020gnv}, SPT \cite{SPT-3G:2022hvq}, BK18 \cite{BICEP:2021xfz} and DESI \cite{DESI:2024mwx} data.


\begin{thebibliography}{99}

\bibitem{Brout:1977ix}
R.~Brout, F.~Englert and E.~Gunzig,
``The Creation of the Universe as a Quantum Phenomenon,'' 
Annals Phys. \textbf{115}, 78 (1978).

\bibitem{Starobinsky:1980te}
A.~A.~Starobinsky,
``A New Type of Isotropic Cosmological Models Without Singularity,''
Phys. Lett. B \textbf{91}, 99-102 (1980).

\bibitem{Kazanas:1980tx}
D.~Kazanas,
``Dynamics of the Universe and Spontaneous Symmetry Breaking,''
Astrophys. J. Lett. \textbf{241}, L59-L63 (1980).

\bibitem{Sato:1981qmu}
K.~Sato,
``First-order phase transition of a vacuum and the expansion of the Universe,''
Mon. Not. Roy. Astron. Soc. \textbf{195}, no.3, 467-479 (1981).

\bibitem{Guth:1980zm}
A.~H.~Guth,
``The Inflationary Universe: A Possible Solution to the Horizon and Flatness Problems,''
Phys. Rev. D \textbf{23}, 347-356 (1981).

\bibitem{Linde:1981mu}
A.~D.~Linde,
``A New Inflationary Universe Scenario: A Possible Solution of the Horizon, Flatness, Homogeneity, Isotropy and Primordial Monopole Problems,''
Phys. Lett. B \textbf{108}, 389-393 (1982).

\bibitem{Linde:1983gd}
A.~D.~Linde,
``Chaotic Inflation,''
Phys. Lett. B \textbf{129}, 177-181 (1983).

\bibitem{Albrecht:1982wi}
A.~Albrecht and P.~J.~Steinhardt,
``Cosmology for Grand Unified Theories with Radiatively Induced Symmetry Breaking,''
Phys. Rev. Lett. \textbf{48}, 1220-1223 (1982).

\bibitem{Martin:2013tda}
J.~Martin, C.~Ringeval and V.~Vennin,
``Encyclop\ae{}dia Inflationaris,''
Phys. Dark Univ. \textbf{5-6} (2014), 75-235.

\bibitem{Achucarro:2022qrl}
A.~Ach\'ucarro \textit{et al.},
``Inflation: Theory and Observations,''
[arXiv:2203.08128 [astro-ph.CO]].

\bibitem{Baumann:2009ds}
D.~Baumann,
``Inflation,''
[arXiv:0907.5424 [hep-th]].

\bibitem{Odintsov:2023weg}
S.~D.~Odintsov, V.~K.~Oikonomou, I.~Giannakoudi, F.~P.~Fronimos and E.~C.~Lymperiadou,
``Recent Advances in Inflation,''
Symmetry \textbf{15}, no.9, 1701 (2023).

\bibitem{Slosar:2019gvt}
A.~Slosar \textit{et al.},
``Scratches from the Past: Inflationary Archaeology through Features in the Power Spectrum of Primordial Fluctuations,''
Bull. Am. Astron. Soc. \textbf{51}, no.3, 98 (2019).

\bibitem{Chluba:2015bqa}
J.~Chluba, J.~Hamann and S.~P.~Patil,
``Features and New Physical Scales in Primordial Observables: Theory and Observation,''
Int. J. Mod. Phys. D \textbf{24}, no.10, 1530023 (2015).

\bibitem{Meerburg:2019qqi}
P.~D.~Meerburg \textit{et al.},
``Primordial Non-Gaussianity,''
Bull. Am. Astron. Soc. \textbf{51}, no.3, 107 (2019).

\bibitem{Maldacena:2002vr}
J.~M.~Maldacena,
``Non-Gaussian features of primordial fluctuations in single field inflationary models,''
JHEP \textbf{05}, 013 (2003)

\bibitem{Chen:2010xka}
X.~Chen,
``Primordial Non-Gaussianities from Inflation Models,''
Adv. Astron. \textbf{2010}, 638979 (2010).

\bibitem{Collins:2011mz}
H.~Collins,
``Primordial non-Gaussianities from inflation,''
[arXiv:1101.1308 [astro-ph.CO]].

\bibitem{Seljak:1996gy}
U.~Seljak and M.~Zaldarriaga,
``Signature of gravity waves in polarization of the microwave background,''
Phys. Rev. Lett. \textbf{78}, 2054-2057 (1997).

\bibitem{Krauss:2010ty}
L.~Krauss, S.~Dodelson and S.~Meyer,
``Primordial Gravitational Waves and Cosmology,''
Science \textbf{328} (2010), 989-992.

\bibitem{Kamionkowski:2015yta}
M.~Kamionkowski and E.~D.~Kovetz,
``The Quest for B Modes from Inflationary Gravitational Waves,''
Ann. Rev. Astron. Astrophys. \textbf{54}, 227-269 (2016).

\bibitem{Caprini:2018mtu}
C.~Caprini and D.~G.~Figueroa,
``Cosmological Backgrounds of Gravitational Waves,''
Class. Quant. Grav. \textbf{35}, no.16, 163001 (2018).

\bibitem{Planck:2019nip}
N.~Aghanim \textit{et al.} [Planck Collaboration],
``Planck 2018 results. V. CMB power spectra and likelihoods,''
Astron. Astrophys. \textbf{641}, A5 (2020).

\bibitem{Planck:2018lbu}
N.~Aghanim \textit{et al.} [Planck Collaboration],
``Planck 2018 results. VIII. Gravitational lensing,''
Astron. Astrophys. \textbf{641}, A8 (2020).

\bibitem{Planck:2018vyg}
N.~Aghanim \textit{et al.} [Planck Collaboration],
``Planck 2018 results. VI. Cosmological parameters,''
Astron. Astrophys. \textbf{641}, A6 (2020)
[erratum: Astron. Astrophys. \textbf{652}, C4 (2021)].

\bibitem{Planck:2018jri}
Y.~Akrami \textit{et al.} [Planck Collaboration],
``Planck 2018 results. X. Constraints on inflation,''
Astron. Astrophys. \textbf{641}, A10 (2020).

\bibitem{Planck:2019kim}
Y.~Akrami \textit{et al.} [Planck Collaboration],
``Planck 2018 results. IX. Constraints on primordial non-Gaussianity,''
Astron. Astrophys. \textbf{641}, A9 (2020).



\bibitem{ACT:2020frw}
S.~K.~Choi \textit{et al.} [ACT Collaboration],
``The Atacama Cosmology Telescope: a measurement of the Cosmic Microwave Background power spectra at 98 and 150 GHz,''
JCAP \textbf{12}, 045 (2020).

\bibitem{ACT:2020gnv}
S.~Aiola \textit{et al.} [ACT Collaboration],
``The Atacama Cosmology Telescope: DR4 Maps and Cosmological Parameters,''
JCAP \textbf{12}, 047 (2020).

\bibitem{SPT-3G:2021eoc}
D.~Dutcher \textit{et al.} [SPT Collaboration],
``Measurements of the E-mode polarization and temperature-E-mode correlation of the CMB from SPT-3G 2018 data,''
Phys. Rev. D \textbf{104}, no.2, 022003 (2021).

\bibitem{SPT-3G:2022hvq}
L.~Balkenhol \textit{et al.} [SPT Collaboration],
``Measurement of the CMB temperature power spectrum and constraints on cosmology from the SPT-3G 2018 TT, TE, and EE dataset,''
Phys. Rev. D \textbf{108}, no.2, 023510 (2023).
	
\bibitem{Wang:2024hhh}
D. Wang,
``Primordial Gravitational Waves 2024,'' 	
[arXiv:2407.02714 [astro-ph.CO]].	
	
\bibitem{BICEP:2021xfz}
P.~A.~R.~Ade \textit{et al.} [BICEP/Keck Collaboration],
``Improved Constraints on Primordial Gravitational Waves using Planck, WMAP, and BICEP/Keck Observations through the 2018 Observing Season,''
Phys. Rev. Lett. \textbf{127}, no.15, 151301 (2021).	

\bibitem{Planck:2015sxf}
P.~A.~R.~Ade \textit{et al.} [Planck Collaboration],
``Planck 2015 results. XX. Constraints on inflation,''
Astron. Astrophys. \textbf{594}, A20 (2016).

\bibitem{Lewis:1999bs}
A.~Lewis, A.~Challinor and A.~Lasenby,
``Efficient computation of CMB anisotropies in closed FRW models,''
Astrophys. J. \textbf{538}, 473-476 (2000).

\bibitem{Lewis:2002ah}
A.~Lewis and S.~Bridle,
``Cosmological parameters from CMB and other data: A Monte Carlo approach,''
Phys. Rev. D \textbf{66}, 103511 (2002).

\bibitem{Lewis:2013hha}
A.~Lewis,
``Efficient sampling of fast and slow cosmological parameters,''
Phys. Rev. D \textbf{87}, no.10, 103529 (2013).

\bibitem{Lewis:2019xzd}
A.~Lewis,
``GetDist: a Python package for analysing Monte Carlo samples,''
[arXiv:1910.13970 [astro-ph.IM]].

\bibitem{Gelman:1992zz}
A.~Gelman and D.~B.~Rubin,
``Inference from Iterative Simulation Using Multiple Sequences,''
Statist. Sci. \textbf{7}, 457-472 (1992).

\bibitem{DESI:2024mwx}
A.~G.~Adame \textit{et al.} [DESI Collaboration],
``DESI 2024 VI: Cosmological Constraints from the Measurements of Baryon Acoustic Oscillations,''
[arXiv:2404.03002 [astro-ph.CO]].

\bibitem{DESI:2024uvr}
A.~G.~Adame \textit{et al.} [DESI Collaboration],
``DESI 2024 III: Baryon Acoustic Oscillations from Galaxies and Quasars,''
[arXiv:2404.03000 [astro-ph.CO]].

\bibitem{DESI:2024lzq}
A.~G.~Adame \textit{et al.} [DESI Collaboration],
``DESI 2024 IV: Baryon Acoustic Oscillations from the Lyman Alpha Forest,''
[arXiv:2404.03001 [astro-ph.CO]].

\bibitem{WMAP:2012nax}
G.~Hinshaw \textit{et al.} [WMAP Collaboration],
``Nine-Year Wilkinson Microwave Anisotropy Probe (WMAP) Observations: Cosmological Parameter Results,''
Astrophys. J. Suppl. \textbf{208}, 19 (2013).

\bibitem{Wang:2024kkk}
D. Wang, In preparation.	

\bibitem{Lyth:1996im}
D.~H.~Lyth,
``What would we learn by detecting a gravitational wave signal in the cosmic microwave background anisotropy?,''
Phys. Rev. Lett. \textbf{78}, 1861-1863 (1997).

\bibitem{Boubekeur:2005zm}
L.~Boubekeur and D.~H.~Lyth,
``Hilltop inflation,''
JCAP \textbf{07}, 010 (2005).

\bibitem{Boubekeur:2012xn}
L.~Boubekeur,
``Theoretical bounds on the tensor-to-scalar ratio in the cosmic microwave background,''
Phys. Rev. D \textbf{87}, no.6, 061301 (2013).

\bibitem{Ooguri:2006in}
H.~Ooguri and C.~Vafa,
``On the Geometry of the String Landscape and the Swampland,''
Nucl. Phys. B \textbf{766}, 21-33 (2007).

\bibitem{Obied:2018sgi}
G.~Obied, H.~Ooguri, L.~Spodyneiko and C.~Vafa,
``De Sitter Space and the Swampland,''
[arXiv:1806.08362 [hep-th]].

\bibitem{Agrawal:2018own}
P.~Agrawal, G.~Obied, P.~J.~Steinhardt and C.~Vafa,
``On the Cosmological Implications of the String Swampland,''
Phys. Lett. B \textbf{784}, 271-276 (2018).

\bibitem{Matsui:2018bsy}
H.~Matsui and F.~Takahashi,
``Eternal Inflation and Swampland Conjectures,''
Phys. Rev. D \textbf{99}, no.2, 023533 (2019).

\bibitem{Ben-Dayan:2018mhe}
I.~Ben-Dayan,
``Draining the Swampland,''
Phys. Rev. D \textbf{99}, no.10, 101301 (2019).

\bibitem{Wang:2018duq}
D.~Wang,
``The multi-feature universe: Large parameter space cosmology and the swampland,''
Phys. Dark Univ. \textbf{28}, 100545 (2020).

\bibitem{Mukhanov:1990me}
V.~F.~Mukhanov, H.~A.~Feldman and R.~H.~Brandenberger,
``Theory of cosmological perturbations. Part 1. Classical perturbations. Part 2. Quantum theory of perturbations. Part 3. Extensions,''
Phys. Rept. \textbf{215}, 203-333 (1992).

\bibitem{Lyth:1998xn}
D.~H.~Lyth and A.~Riotto,
``Particle physics models of inflation and the cosmological density perturbation,''
Phys. Rept. \textbf{314}, 1-146 (1999).

\bibitem{Copeland:1994vg}
E.~J.~Copeland, A.~R.~Liddle, D.~H.~Lyth, E.~D.~Stewart and D.~Wands,
``False vacuum inflation with Einstein gravity,''
Phys. Rev. D \textbf{49}, 6410-6433 (1994).

\bibitem{Dine:1995uk}
M.~Dine, L.~Randall and S.~D.~Thomas,
``Supersymmetry breaking in the early universe,''
Phys. Rev. Lett. \textbf{75}, 398-401 (1995).

\bibitem{Kundu:2013gha}
S.~Kundu,
``Non-Gaussianity Consistency Relations, Initial States and Back-reaction,''
JCAP \textbf{04}, 016 (2014).

\bibitem{Verbiest:2021kmt}
J.~P.~W.~Verbiest, S.~Oslowski and S.~Burke-Spolaor,
``Pulsar Timing Array Experiments,''
[arXiv:2101.10081 [astro-ph.IM]].

\bibitem{LIGOScientific:2016jlg}
B.~P.~Abbott \textit{et al.} [LIGO-Virgo Collaboration],
``Upper Limits on the Stochastic Gravitational-Wave Background from Advanced LIGO\textquoteright{}s First Observing Run,''
Phys. Rev. Lett. \textbf{118}, no.12, 121101 (2017)
[erratum: Phys. Rev. Lett. \textbf{119}, no.2, 029901 (2017)].

\bibitem{NANOGrav:2023gor}
G.~Agazie \textit{et al.} [NANOGrav Collaboration],
``The NANOGrav 15 yr Data Set: Evidence for a Gravitational-wave Background,''
Astrophys. J. Lett. \textbf{951}, L8 (2023).

\bibitem{Antoniadis:2023rey}
J.~Antoniadis \textit{et al.} [EPTA Collaboration],
``The second data release from the European Pulsar Timing Array III. Search for gravitational wave signals,''
[arXiv:2306.16214 [astro-ph.HE]].

\bibitem{Reardon:2023gzh}
D.~J.~Reardon \textit{et al.} [PPTA Collaboration],
``Search for an Isotropic Gravitational-wave Background with the Parkes Pulsar Timing Array,''
Astrophys. J. Lett. \textbf{951}, L6 (2023)

\bibitem{Xu:2023wog}
H.~Xu \textit{et al.} [CPTA Collaboration],
``Searching for the Nano-Hertz Stochastic Gravitational Wave Background with the Chinese Pulsar Timing Array Data Release I,''
Res. Astron. Astrophys. \textbf{23}, 075024 (2023).

\bibitem{NANOGrav:2023hvm}
A.~Afzal \textit{et al.} [NANOGrav Collaboration],
``The NANOGrav 15 yr Data Set: Search for Signals from New Physics,''
Astrophys. J. Lett. \textbf{951}, L11 (2023).

\bibitem{Antoniadis:2022pcn}
J.~Antoniadis, \textit{et al.} [IPTA Collaboration],
``The International Pulsar Timing Array second data release: Search for an isotropic gravitational wave background,''
Mon. Not. Roy. Astron. Soc. \textbf{510}, no.4, 4873-4887 (2022).

\bibitem{Guzzetti:2016mkm}
M.~C.~Guzzetti, N.~Bartolo, M.~Liguori and S.~Matarrese,
``Gravitational waves from inflation,''
Riv. Nuovo Cim. \textbf{39}, no.9, 399-495 (2016).


















\end{thebibliography}
\end{document}